\input harvmac
\def\MM{{\cal M}}
\Title{\vbox{\baselineskip12pt
\hbox{hep-th/9712146}
\hbox{DAMTP/97-145 }
\hbox{EFI-97-57 }
}}
 {\vbox{\centerline{Monstrous Heterotic Quantum Mechanics}}}
\medskip
\centerline{\it Michael B.  
Green\footnote{*}{M.B.Green@damtp.cam.ac.uk}}
\centerline{DAMTP, Silver Street, Cambridge CB3 9EW, UK}
\vskip .2in 
\centerline{\it David 
Kutasov\footnote{**}{kutasov@yukawa.uchicago.edu}}
\smallskip
\centerline{EFI and Department of Physics, University of Chicago}
\centerline{5640 South Ellis Avenue, Chicago, IL 60637, USA}
\centerline{and}
\centerline{Department of Physics of Elementary Particles}
\centerline{Weizmann Institute of Science}
\centerline{Rehovot, 76100, Israel}
\vglue .3cm
\bigskip
 
\noindent
A compactification of the heterotic 
string is considered which describes 
a particle propagating in the symmetric space 
$\MM=SO(9,1;Z)\backslash SO(9,1;R)/SO(9;R)$
in interaction with excited states whose dynamics is
$1+1$ dimensional. The model appears to describe 
D-particle dynamics in type IIA string theory on 
$\MM$.

\Date{12/97}

\def\journal#1&#2(#3){\unskip, \sl #1\ \bf #2 \rm(19#3) }
\def\andjournal#1&#2(#3){\sl #1~\bf #2 \rm (19#3) }

\def\ie{{\it i.e.}}
\def\eg{{\it e.g.}}

\def\frac#1#2{{#1\over#2}}

\def\inbar{\,\vrule height1.5ex width.4pt depth0pt}
\def\IC{\relax\hbox{$\inbar\kern-.3em{\rm C}$}}
\def\IR{\relax{\rm I\kern-.18em R}}
\def\IP{\relax{\rm I\kern-.18em P}}

%
%
\def\np#1#2#3{Nucl. Phys. {\bf B#1} (#2) #3}
\def\pl#1#2#3{Phys. Lett. {\bf #1B} (#2) #3}

\def\prl#1#2#3{Phys. Rev. Lett. {\bf #1} (#2) #3}

\def\prd#1#2#3{Phys. Rev. {\bf D#1} (#2) #3}

\def\cqg#1#2#3{Class. Quant. Grav. {\bf #1} (#2) #3}

\catcode`\@=11
\def\slash#1{\mathord{\mathpalette\c@ncel{#1}}}
\overfullrule=0pt

\def\LL{{\cal L}}
\def\MM{{\cal M}}

\def\underrel#1\over#2{\mathrel{\mathop{\kern\z@#1}\limits_{#2}}}

\catcode`\@=12


%

\def\exp{{\rm exp}}



\nref\cremjul{E. Cremmer and B. Julia, 
{\it The $N=8$ Supergravity Theory. 1.  The Lagrangian}, 
  \pl{80}{1978}{48};  {\it The $SO(8)$ Supergravity},  
\np{159}{1979}{141}.} 
\nref\hulltown{C. M. Hull and P. K. Townsend, 
{\it Unity of Superstring Dualities}, hep-th/9410167,
\np{438}{1995}{109}.}
\nref\witt{E. Witten, 
{\it String Theory Dynamics in Various Dimensions},  hep-th/9503124, 
\np{443}{1995}{85}.}
\nref\egkr{S. Elitzur, A. Giveon, D. Kutasov and 
E. Rabinovici, {\it Algebraic Aspects of Matrix 
Theory on $T^D$}, hep-th/9707217.}
\nref\hull{C. M. Hull, {\it Matrix Theory, 
U-duality and Toroidal Compactifixcations of M-theory},  
hep-th/9711179; {\it U-duality and BPS Spectrum of Super 
Yang--Mills Theory and M-Theory},  hep-th/9712075.}
\nref\blo{M. Blau and M. O'Loughlin, {\it Aspects of 
U-duality in Matrix Theory}, hep-th/9712047.}
\nref\elie{N. A. Obers, B. Pioline and E. Rabinovici, 
{\it M-Theory and U-Duality on $T^D$ with Gauge Backgrounds}, 
hep-th/9712084.}
\nref\bs{T. Banks and L. Susskind,
{\it The Number of States 
of Two-Dimensional Critical String Theory}, hep-th/9511193,
\prd{54}{1996}{1677}.} 
\nref\green{M. B. Green, {\it World-Sheets for World-Sheets},  
\np{293}{1987}{593}.}
\nref\km{D. Kutasov and E. Martinec,
{\it New Principles for String/Membrane Unification},  hep-th/9602049,
\np{477}{1996}{652}; 
{\it M-branes and $N=2$ Strings}, hep-th/9612102, \cqg{14}
{1997}{2483}; D. Kutasov, E. Martinec and M. 
O'Loughlin, {\it Vacua of M-theory and $N=2$ Strings}, 
hep-th/9603116, \np{477}{1996}{675}.} 
\nref\bgr{L. Baulieu, M. B. Green and E. Rabinovici, 
{\it Superstrings from Theories with $N>1$ Supersymmetry},
hep-th/9611136, \np{498}{1997}{119}.}
\nref\dga{ L. Dixon,  P. Ginsparg,  J.A. Harvey,  {\it Beauty and the Beast: Superconformal Symmetry in a Monster Module},   Commun. Math. Phys. {\bf 119}  (1988)  221.}
\nref\dgm{L.  Dolan, P.  Goddard and P.  Montagu, {\it Conformal Field Theory of Twisted Vertex Operators}, Nucl.  Phys.  {\bf B338} (1990) 529.}
 \nref\cont{D. Kutasov, {\it Geometry of the Space of 
Conformal Field Theory and Contact Terms}, 
\pl{220}{1989}{153}.}
\nref\dabh{A. Dabholkar and J. Harvey,  
{\it Nonrenormalization of the Superstring Tension}, 
\prl{63}{1989}{478}.}

Low dimensional compactifications of string/M-theory
are of great interest for deciphering the underlying
symmetries of the theory. The non-perturbative 
U-duality symmetry group becomes ever larger as the 
number of non-compact dimensions is reduced
\refs{\cremjul-\witt}. There are arguments 
(see \eg\ \refs{\egkr-\elie} for recent discussions and
further references) that suggest that the symmetry
group of two dimensional type II string theory 
(or M-theory on a nine-torus) is a discrete version
of the affine group $\hat E_8$ (sometimes refered to
as $E_9$), while in one non-compact dimension (time)
the symmetry group is a discrete hyperbolic group
known as $E_{10}$. 
At the same time, there are many conceptual problems
with defining static vacua in low dimensional theories
of gravity (see \eg\ \bs).

Another aspect of low dimensional string theory
is the observation of \green\ further explored
in \km\ that low dimensional string theory may
play an important role in describing in a unified
way the worldvolume geometry of critical strings
and other branes. This circle of ideas has not yet
found its proper place in the emerging understanding
of non-perturbative string theory, but it is believed
that string theories with extended world-sheet SUSY
may be of special significance in this regard
\refs{\green-\bgr}. 

In this note we discuss the properties of an 
unusual compactification of the standard heterotic
string   to one space-time dimension, which
preserves all sixteen supercharges of the original
ten dimensional theory\foot{This theory was incompletely
discussed in \green.}. 
Normally, such systems have a Narain moduli space
of classical vacua. In $0+1$ dimensions,
the low energy dynamics is instead described by a 
$\sigma$-model for the light fields whose target
space is the classical moduli space. Clearly, to 
understand it we have to keep all the string
modes that become light anywhere in moduli space.
In different decompactification regions of the 
classical moduli space the spectrum of light
states becomes continuous, revealing the higher
dimensional nature of the theory. 

\noindent
The example we will discuss has two special
features:
\item{(1)} The low energy dynamics corresponds
to quantum mechanics with a finite number of 
degrees of freedom everywhere in the moduli space
of vacua. There is a finite energy gap
that separates the low energy quantum mechanics  from the
field theoretic continuum.

\item{(2)} While the original heterotic string
lives on $R\times T^9$, its space-time dynamics
describes particle propagation on a different
manifold $R\times \MM$ which is curved and has
a cusp. 

We start with the standard compactification 
of the heterotic string on $T^9$. The left
and right moving coordinates of the string
$(\bar x^1,\cdots, \bar x^{25}; x^1,\cdots, x^9)$
live on an even self-dual Lorentzian torus 
$\Gamma^{25,9}$; the $25\times9$ dimensional
Narain moduli space of inequivalent tori
\eqn\mhet{
\MM_{\rm Narain}=SO(25,9;Z)\backslash SO(25,9;R)/SO(25;R)\times
SO(9;R),} 
is parametrized by expectation values of the scalar
fields
\eqn\modli{V_{ia}=\partial x^i\bar\partial\bar x^a
e^{iEx^0}}
where $i=1,\cdots, 9$, $a=1,\cdots, 25$, and
$E=0$ on shell. In the limit of decompactification to ten dimensions 
the torus takes the form,
\eqn\decm{\Gamma^{25,9}=\Gamma^{16}\oplus\Gamma^{1,1}_{R_1}
\oplus\Gamma^{1,1}_{R_2}\oplus\cdots\oplus\Gamma^{1,1}_{R_9},}
where $\Gamma^{16}$ is one of the two 
sixteen dimensional even self-dual Euclidean tori and the
nine radii $R_i$ labeling the $\Gamma^{1,1}$'s
are taken to infinity. In this region of moduli space,
the heterotic quantum mechanics approaches $9+1$
dimensional field theory. 

However, there is an interesting projection of the 
compactified theory onto 
a subspace that defines a theory that has no continuum 
states and no field theory limit.  This is the subspace 
of $\MM_{\rm Narain}$ in which the Lorentzian lattice 
takes the form,
\eqn\niem{\Gamma^{25,9}=\Gamma^{24}\oplus\Gamma^{1,9},}
where $\Gamma^{24}$ is the Leech torus, which is the unique 
twenty-four dimensional  even   self-dual  torus that 
has no vectors of length $\sqrt2$.  Since $\Gamma^{24}$ has no 
moduli   all the moduli of $\Gamma^{25,9}$ are contained in the  
Lorentzian even self-dual
torus, $\Gamma^{1,9}$, that has  a nine-dimensional moduli 
space,
\eqn\mnone{ 
\MM=SO(9,1;Z)\backslash SO(9,1;R)/SO(9;R).}
  The projection 
of interest is obtained by performing
an asymmetric orbifold twist on the left-moving
Leech torus,
\eqn\twmon{(\bar x^2, \bar x^3,\cdots, \bar x^{25})
\rightarrow - (\bar x^2, \bar x^3,\cdots, \bar x^{25}).}
The $c=24$ left-moving CFT corresponding to 
\twmon\ is the Monster module \refs{\dga,\dgm}. 
The torus partition sum of the resulting heterotic
vacuum is
\eqn\partmon{Z(\tau, \bar\tau)={1\over2}
\left[\left({\theta_3(\tau)\over\eta(\tau)}\right)^4-
\left({\theta_4(\tau)\over\eta(\tau)}\right)^4-
\left({\theta_2(\tau)\over\eta(\tau)}\right)^4\right]
J(\bar\tau){\sqrt{\tau_2}\over\eta(\tau)^8}
\sum_{(p_R, p_L)\in \Gamma^{9,1}}
q^{{1\over2}p_R^2}\bar q^{{1\over2}p_L^2},}
where $\theta_i$ are Jacobi $\theta$ functions,
$J$ is the unique meromorphic modular invariant
with vanishing constant term,
\eqn\Jt{J(\bar\tau)={1\over \bar q}+0+C\bar q+\cdots,}
and $q=\exp(2\pi i\tau)$. The fact that the 
constant term in \Jt\ vanishes means that
there are no dimension one operators in the monster
module CFT. Furthermore, it is clear from 
\partmon\ that the vacuum preserves the same
sixteen supercharges as the original heterotic
model on $T^9$ and the classical moduli space
of vacua is $\MM$ \mnone. 

The light degrees of freedom of the theory are as
follows. At generic points in $\MM$ only the nine
moduli fields $\Phi_i$ $(i=1,\cdots, 9)$, with
vertex operators,
\eqn\vmod{V_i=\partial x^i\bar\partial \bar x^1 e^{iEt},}
and their sixteen fermionic partners have zero energy
on shell. As we approach the point in $\MM$ where
\eqn\genpt{
\Gamma^{9,1}=\Gamma^8\oplus\Gamma^{1,1}_{R=\sqrt2},}
where $\Gamma^8$ is the $E_8$ torus  
and $\Gamma^{1,1}_{R=\sqrt2}$ corresponds 
to a compactification on a circle of radius
$R=\sqrt2$, additional light states appear from
vectors with $p_L^2=2$, $p_R=0$ in \partmon.
These states $\Phi_i^\pm$ are described by the vertex
operators 
\eqn\chst{V_i^\pm=\partial x^i e^{\pm i\sqrt2
\bar x^1} e^{iEt}}
and, together with $\Phi_i\equiv \Phi_i^3$, fill out
an adjoint multiplet $\Phi_i^a$ 
$(a=3, \pm$; $i=1,\cdots, 9)$ of the $SU(2)$ gauge
symmetry generated by $\bar\partial \bar x^1$,
$\exp(\pm i\sqrt{2}\bar x^1)$. 

The low energy Lagrangian describing $\Phi_i^a(t)$
locally near a degeneration point \genpt\ is the
dimensional reduction of $N=1$ SYM with gauge group
$SU(2)$ in $9+1$ dimensions   to $0+1$ dimensions.
Globally, the manifold $\MM$ is curved and has a cusp
corresponding to \genpt. Obviously, there are images
of this cusp obtained by acting with $SO(9,1;Z)$. 
However, there are no inequivalent cusps, such as  ones
at which a larger number of states become 
massless simultaneously. The low energy Lagrangian
for the modes $\Phi$ \vmod\  generically
has the form, 
\eqn\lgij{\LL=G_{ij}(\Phi)\dot\Phi^i\dot\Phi^j,}
where $G_{ij}$ is the Zamolodchikov metric
on the symmetric space \mnone. This can be verified
by standard techniques \refs{\cont}. Near the cusp one must
add the degrees of freedom \chst\ to obtain a 
non-singular description.

The system in question has a large number of
$1/2$ BPS states obtained as in \refs{\dabh} 
 by putting the
right-movers in their ground state and allowing
arbitrary excitations (consistent with the
projection \twmon) of the left-movers. 
The level matching condition is
\eqn\levmat{1+{1\over2}p_R^2={1\over2}
p_L^2+N_L}
and the energy of the corresponding
states is
\eqn\EBPS{E=|\vec p_R|.}
Restricting to the region of moduli space
in which $\Gamma^{9,1}=\Gamma^8
\oplus\Gamma^{1,1}_R$ and taking the
decompactification limit $R\to\infty$
gives  $|\vec p_R|^2=k^2+|\vec p_{E_8}|^2$,
where $k$ is the continuous momentum in the
non-compact $x^1$ direction and $\vec p_{E_8}
\in \Gamma^8$. We see, therefore, that these BPS states form 
a continuum as $R\to\infty$, \ie\ they are described
by a $1+1$ dimensional field theory. However,
they are separated from the low energy
quantum mechanics by a gap. States with $\vec p_{E_8}\not
=0$ have energy $\ge |\vec p_{E_8}|$. 
States with $\vec p_{E_8}=0$ cannot be massless
since the monster CFT does not have dimension
one operators.  
We conclude that the two dimensional field 
theoretic continuum is separated from the low
energy quantum mechanics by an energy gap of order one in string 
units. 

In view of string duality one can ask whether
the vacuum described above has an alternative
description. A natural candidate is a `compactification'
of type IIA string theory on the manifold $\MM$ 
\mnone\ in the presence of a D-particle.
When the D-particle approaches the cusp of $\MM$ 
\genpt, states corresponding to fundamental strings
stretched between the  particle and its image
with respect to the $Z_2$ fixed point become light
and enhance the $U(1)$  quantum mechanics to $SU(2)$. 

Such a picture would provide a natural explanation
of the $SU(2)$ quantum mechanics discussed above, as well as of the 
energy gap to the excited states. It would be an interesting
example of consistent D-particle propagation on a curved
space.

More work is necessary to establish whether type IIA strings
can be consistently formulated on $\MM\times R$ in the
presence of a D-particle. It is possible that one will
also need to place an orientifold at the cusp of $\MM$ 
in order to satisfy the equations of motion. If this is the case,
the charge of the orientifold plane would have to be positive
to get an $Sp(1)\simeq SU(2)$ projection on a nearby
D-particle (negative charge would instead give rise
to $SO(2)$). It would be interesting to clarify this issue
and show that such a vacuum has the right amount of SUSY.

The heterotic vacuum we have discussed seems also
to be related to the supersymmetric $(2,1)$ 
heterotic string vacuum studied in \km. In particular,
the two have the same amount of supersymmetry
and moduli space of vacua $\MM$ \mnone. 
Thus, a better understanding
of this theory may provide useful insights for
the program of \refs{\green, \km}. 

\bigskip
\noindent{\bf Acknowledgements:}
We thank E. Martinec for useful discussions
and the Aspen Center for Physics for hospitality.
The work of D. K. is supported in part by a DOE OJI
grant.

\listrefs

\bye